\documentclass[lettersize,journal]{IEEEtran}
\usepackage{amsmath,amsfonts,amsthm}
\usepackage{algorithmic}
\usepackage{algorithm}
\usepackage{array}
\usepackage{textcomp}
\usepackage{url}
\usepackage{verbatim}
\usepackage{graphicx}
\usepackage{makecell}
\usepackage{multirow}
\usepackage{booktabs}
\usepackage{subcaption}
\usepackage{cite}
\usepackage{braket}

\newtheorem{proposition}{Proposition}

\newcommand{\DO}{\mathcal{DO}}

\newcommand{\Tr}{\mathrm{Tr}}

\begin{document}

\title{Implementing Pearl's $\mathcal{DO}$-Calculus on Quantum Circuits: A Simpson-Type Case Study on NISQ Hardware}

\author{Pilsung Kang
\thanks{Pilsung Kang is with the Department of Software Science, Dankook University, Yongin 16890, South Korea (e-mail: pilsungk@dankook.ac.kr)}
}

\maketitle

\begin{abstract}
Distinguishing correlation from causation is a central challenge in machine intelligence, and Pearl's $\mathcal{DO}$-calculus provides a rigorous symbolic framework for reasoning about interventions. A complementary question is whether such intervention logic can be given \emph{executable semantics} on physical quantum devices. Our approach maps causal networks onto quantum circuits, where nodes are encoded in qubit registers, probabilistic links are implemented by controlled-rotation gates, and interventions are realized by a structural remodeling of the circuit---a physical analogue of Pearl's ``graph surgery'' that we term \emph{circuit surgery}. We show that, for a family of 3-node confounded treatment models (including a Simpson-type reversal), the post-surgery circuits reproduce exactly the interventional distributions prescribed by the corresponding classical $\mathcal{DO}$-calculus. We then demonstrate a proof-of-principle experimental realization on an IonQ Aria trapped-ion processor and a 10-qubit synthetic healthcare model, observing close agreement between hardware estimates and classical baselines under realistic noise. We do not claim quantum speedup; instead, our contribution is to establish a concrete pathway by which causal graphs and Pearl-style interventions can be represented, executed, and empirically tested within the formalism of quantum circuits.
\end{abstract}

\begin{IEEEkeywords}
Quantum causal inference, Quantum machine learning, Pearl's $\mathcal{DO}$-calculus, Circuit surgery, Quantum circuits, Simpson's paradox, NISQ devices
\end{IEEEkeywords}

%%%%%%%%%%%%%%%%%%%%%%%%%%%%%%%%%%%%%%%%%%%%%%%%%%%%%%%%%%%%%%%%%%%%%%%% 
\section{Introduction}\label{s:intro}

Distinguishing correlation from genuine cause--effect relationships is a foundational problem in statistics and machine learning, and is increasingly important for data-driven AI systems deployed in the real world~\cite{pearl:2009:causality}. This distinction is not merely academic; in high-stakes domains such as medical diagnostics~\cite{ueda:2024}, financial modeling~\cite{kumar:2025:csur}, and autonomous systems~\cite{giamattei:2024:tsem}, decisions based on spurious correlations can lead to unsafe, biased, and unreliable outcomes. Consequently, moving beyond pattern recognition to build models that can reason about cause and effect is a central challenge for developing robust, fair, and trustworthy AI systems.

A classic illustration of this challenge is Simpson's Paradox~\cite{simpson:1951,pearl:2022:understanding}, a statistical phenomenon where a trend observed across an entire population reverses within its constituent subgroups once the data is partitioned. This phenomenon is particularly pernicious in the causal analysis of observational data, for example, when evaluating the efficacy of a medical treatment on a patient's outcome. In such scenarios, the paradox emerges when a hidden confounding variable---such as patient demographics that act as a common cause of both the treatment decision and the health outcome---is unevenly distributed across the subgroups. A machine learning model trained on such aggregated data without knowledge of the underlying causal structure would learn the spurious, reversed correlation, leading to fundamentally flawed conclusions. From a causal-inference perspective, Simpson's Paradox thus serves as a canonical stress test: any framework that aspires to compute interventional quantities must explain how such reversals arise from confounding and how they are resolved by appropriate adjustment.

The theoretical foundation for analyzing such situations and, more broadly, for reasoning about causal effects is provided by Pearl's $\mathcal{DO}$-calculus~\cite{pearl:1995:do}. The $\mathcal{DO}$-calculus offers a formal framework for computing the results of interventions, such as the interventional probability $P(Y \mid \mathcal{DO}(X))$, from observational data when the underlying causal graph is known. This framework has become a reference point in modern causal inference, allowing researchers to systematically distinguish causal effects from spurious associations. In classical causal inference, this calculus underpins a broad toolbox of identification and estimation techniques that are routinely applied to observational data. At the same time, it is typically realized as an abstract, symbolic procedure on probability distributions rather than as a concrete physical process. A natural question, therefore, is whether this calculus of interventions can also be instantiated in the dynamics of a physical system, providing an alternative computational paradigm for causal reasoning.

Quantum computing presents a natural candidate for such a physical realization, owing to a useful structural analogy between structural causal models (SCMs) and quantum circuits: both are probabilistic at the level of observed outcomes, directed, and defined by local interactions. This analogy motivates the question of how the abstract ``graph surgery'' of the $\mathcal{DO}$-calculus can be mapped to concrete operations on a quantum circuit.

In prior work~\cite{kang:2025:esc-arxiv}, we introduced a framework for quantum causal inference, establishing how quantum circuits can model causal interventions based on Pearl's $\mathcal{DO}$-calculus. In this paper, we apply and extend this framework to a canonical challenge in causal reasoning: Simpson's Paradox. We validate our approach through a series of simulations and a proof-of-principle experiment on quantum hardware. Our contributions are threefold:
\begin{itemize}
  \item We compile binary SCMs into block-diagonal quantum circuits, representing nodes as qubit registers, encoding conditional probabilities via controlled-rotation gates, and giving Pearl's $\mathcal{DO}$-operator executable semantics through a simple circuit-surgery rule.
  \item We instantiate this construction on a 3-qubit confounded treatment model that exhibits a Simpson-type reversal, and validate the $\mathcal{DO}$-calculus resolution both in high-fidelity simulation and on an IonQ Aria trapped-ion noisy intermediate-scale quantum (NISQ) processor, observing close agreement with classical baselines under realistic hardware noise.
  \item We apply the same methodology to a 10-qubit synthetic healthcare network with multi-level confounding, using the circuit-level implementation of backdoor adjustment to quantify confounding bias beyond what simple statistical stratification can correct.
\end{itemize}
Throughout, we make no claim of quantum advantage or new causal identification criteria; instead, we offer an experimental validation of classical $\mathcal{DO}$-calculus on current NISQ hardware and a concrete case study of how SCM-style graph surgery can be executed and empirically tested within the formalism of quantum circuits.

The remainder of this paper is organized as follows. Section~\ref{s:related} reviews related work on quantum causal models, Born-rule generative models, and quantum analogues of Simpson effects. Section~\ref{s:method} introduces our causal framework, the circuit-surgery implementation of $\mathcal{DO}$-operations, and the specific 3-qubit and 10-qubit models used in this study. Section~\ref{s:results} presents simulation and hardware results, comparing ideal and noisy implementations and quantifying confounding bias. Section~\ref{s:discuss} discusses the interpretation of these findings, connections to classical methods, and current limitations and future directions. Finally, Section~\ref{s:conc} concludes the paper.

%%%%%%%%%%%%%%%%%%%%%%%%%%%%%%%%%%%%%%%%%%%%%%%%
\section{Related Work}\label{s:related}

Classical causal inference provides a mature toolbox for identifying and estimating causal effects from observational data, grounded in structural causal models and the $\mathcal{DO}$-calculus~\cite{pearl:2009:causality,pearl:1995:do}. Constraint-based discovery algorithms such as the Fast Causal Inference (FCI) procedure~\cite{spirtes:2000:book}, functional-model approaches like LiNGAM for linear non-Gaussian systems~\cite{shimizu:2006:lingam}, and identification strategies based on covariate adjustment, instrumental variables, and the front-door criterion~\cite{pearl:2009:causality} offer well-established ways to handle confounding and to recover quantities such as $P(Y \mid \mathcal{DO}(X))$ in fully classical settings. Our goal in this work is not to propose an alternative to these methods, but to investigate how the same intervention logic can be compiled into and experimentally validated on quantum circuits.

In what follows, we briefly review three strands: (i) quantum causal-model frameworks (conditional states, process matrices, and quantum combs), (ii) Born-rule-based quantum-circuit generative models (``Born machines''), and (iii) quantum analogues of Simpson effects driven by interference and contextuality. We then clarify how our approach differs by compiling $\mathcal{DO}$-operations into circuits and validating them on NISQ hardware.

\subsection{Quantum Causal Modeling}
A mature line of work generalizes classical causal models to quantum systems by replacing random variables with quantum systems and conditional probabilities with completely positive (CP) maps or ``conditional states.'' Foundational formalisms include the conditional-state approach of Leifer and Spekkens (quantum Bayesian inference)~\cite{leifer:2013}, the quantum-combs framework that represents multi-time processes and interventions as higher-order maps~\cite{chiribella:2009}, and the process-matrix formalism that allows for indefinite causal order~\cite{oreshkov:2012}. On top of these primitives, quantum causal models (QCMs) and discovery algorithms have been proposed, aiming to recover causal structure from data and to define quantum analogues of interventions~\cite{costa:2016:qcm,barrett:2020:qcm,giarmatzi:2018:qcda}.

Conceptually, QCMs separate causal structure (wiring of CP instruments) from dynamics (the CP maps themselves), mirroring the classical split between a directed acyclic graph (DAG) and its structural equations~\cite{barrett:2020:qcm}. In definite-order (acyclic) settings, interventions correspond to replacing instruments along a node (quantum $\mathcal{DO}$ operations), whereas the process-matrix line shows that some quantum processes are causally nonseparable and thus lie beyond any single underlying order~\cite{oreshkov:2012}. Our work is complementary: we adopt the same operator-theoretic primitives (states, effects, and instruments) to give executable semantics to classical $\mathcal{DO}$-operations, but we do not build a full QCM nor engage with indefinite order; instead, we target a concrete causal task (Simpson's Paradox) and validate interventional predictions on NISQ hardware with direct comparability to classical baselines.

\subsection{Born-rule Generative Models (Born Machines)}

A parallel line of work leverages the Born rule~\cite{nielsen2010quantum} to define generative models, where a parameterized quantum circuit assigns probabilities to classical data via measurement outcomes. Early formulations—often called Quantum Circuit Born Machines (QCBMs)—optimize circuit parameters so that the Born distribution matches a target dataset, using objectives such as maximum mean discrepancy, likelihood surrogates, or adversarial criteria~\cite{liu:2018:qcbm,benedetti:2019:pqc}. Subsequent studies explored scaling, expressivity, and potential advantage of Ising-type Born machines for sampling tasks~\cite{coyle:2020:born}, while also identifying trainability challenges in variational circuits (barren plateaus)~\cite{mcclean:2018:barren}.

Our use of the Born rule is orthogonal to these generative approaches. We do not train a parameterized circuit to fit a dataset; rather, we use the operator calculus to compile classical $\mathcal{DO}$-operations (intervene, condition, and marginalize) into circuit-level primitives (state preparation, controlled gates, and measurements) and to validate that the resulting interventional probabilities resolve Simpson's Paradox exactly as predicted by SCM. Phase serves only as a controllable degree of freedom to reproduce classical interaction contrasts.

\subsection{Quantum Analogues of Simpson's Paradox}

Several works analyze ``quantum Simpson effects,'' where aggregation reversals arise from noncommuting measurements and amplitude interference rather than classical confounding~\cite{shi:2012:qs,paris:2012:twoqs}. These studies sharpen the conceptual distinction between interference-driven and confounder-driven reversals. Our focus is different: we operationalize the classical causal resolution by compiling $\mathcal{DO}$-operations into circuits and verifying them on hardware, thereby providing a physically grounded implementation of interventional logic rather than a contextuality-based account.

%%%%%%%%%%%%%%%%%%%%%%%%%%%%%%%%%%%%%%%%%%%%%%%%%%%%
\section{Methodology}\label{s:method}

In this section we specify the Simpson-type SCMs we study, recall the relevant $\mathcal{DO}$-calculus notions, and describe how these ingredients are compiled into quantum circuits via circuit surgery.
For completeness, the formal SCM-to-circuit notation underlying this construction is summarized in Appendix B.

\subsection{Causal Framework and the $\mathcal{DO}$-calculus}

Simpson's Paradox is a statistical phenomenon wherein a trend that appears in several groups of data disappears or reverses when these groups are combined. This effect is particularly common when analyzing observational data to evaluate, for instance, the efficacy of a medical \textit{treatment} on a patient's \textit{outcome}. Within the framework of Pearl's structural causal models (SCMs), the paradox is understood not as a mathematical anomaly but as a consequence of misinterpreting these causal relationships. Specifically, it often arises when a hidden confounding variable influences both the treatment and the outcome, creating a spurious correlation that can obscure the true causal effect.

\begin{figure}[htbp]
    \centering
    % (a) Observational DAG
    \begin{subfigure}[b]{0.45\textwidth}
        \centering
        \includegraphics[width=0.7\textwidth]{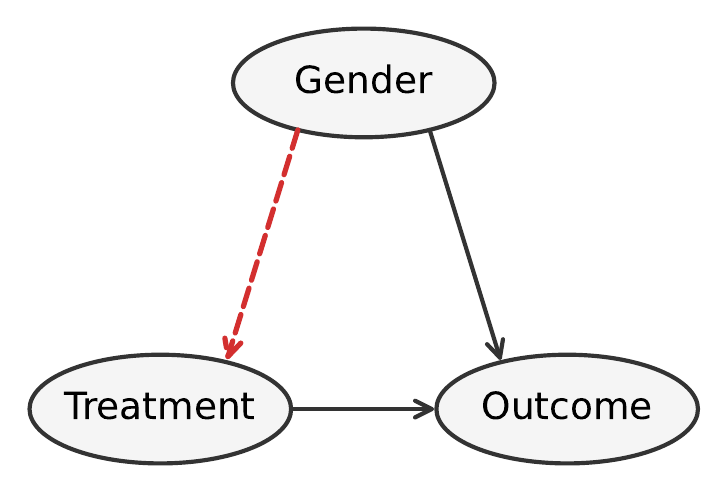}
        \caption{Observational Causal Graph}
        \label{fig:dag_obs}
    \end{subfigure}
    \hfill 
    % (b) Interventional DAG
    \begin{subfigure}[b]{0.45\textwidth}
        \centering
        \includegraphics[width=0.7\textwidth]{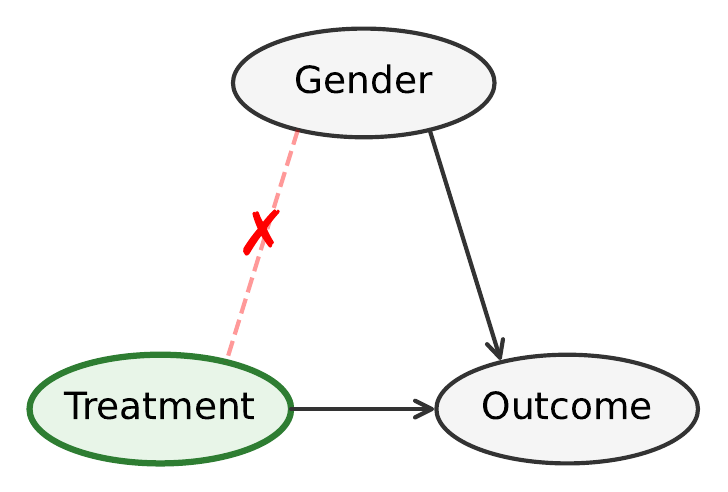}
        \caption{Interventional Causal Graph}
        \label{fig:dag_int}
    \end{subfigure}
    \caption{The causal directed acyclic graphs (DAGs) for the 3-qubit Simpson's Paradox model. (a) The observational model, where G (Gender) is a common-cause confounder for T (Treatment) and O (Outcome). (b) The interventional model for $\mathcal{DO}(T=t)$, where the confounding path $G \rightarrow T$ has been surgically removed.}
    \label{fig:causal_dags}
\end{figure}

In the scenario modeled in this work, the paradox is driven by a confounder, Gender (G), which affects both the Treatment (T) administered and the final Outcome (O).  This causal structure can be represented by a DAG, as shown in Fig.~\ref{fig:dag_obs}.  The confounder G has a direct causal influence on both T ($G \rightarrow T$) and O ($G \rightarrow O$), while T also has a direct influence on O ($T \rightarrow O$).  The path $G \rightarrow T$ represents the confounding bias, such as a propensity for one gender to receive a particular treatment more often than the other.  To resolve the paradox, we must distinguish between two types of conditional probabilities:

%% \begin{figure}[htbp]
%%     \centering
%%     \includegraphics[width=0.7\columnwidth]{./Images/causal_dag_3q.pdf}
%%     \caption{The causal Directed Acyclic Graph (DAG) for the 3-qubit Simpson's Paradox model. The variable Gender (G) acts as a common-cause confounder, influencing both the Treatment (T) and the Outcome (O). The arrow from T to O represents the direct causal effect we aim to measure.}
%%     \label{fig:causal_dag}
%% \end{figure}
%% 
% ------------------------------------

\begin{enumerate}
    \item \textbf{Observational Probability, $P(O|T)$:} This corresponds to the passive observation of ``the probability of a good Outcome, given that we \textit{see} a subject has taken the Treatment.'' This probability incorporates not only the direct causal effect of T on O but also the spurious correlation created by the confounding path through G. Simpson's Paradox manifests in this observational distribution.

    \item \textbf{Interventional Probability, $P(O|\mathcal{DO}(T))$:} This corresponds to the question ``what is the probability of a good Outcome \textit{if we force} all subjects to take the Treatment?'', which is equivalent to the outcome of a randomized controlled trial (RCT). In the causal graph, the $\mathcal{DO}(T)$ operator is realized by performing ``graph surgery'': all incoming arrows to the variable T (in this case, the $G \rightarrow T$ link) are severed, resulting in the modified causal graph shown in Fig.~\ref{fig:dag_int}. By blocking the confounding path, $P(O|\mathcal{DO}(T))$ isolates the pure causal effect of T on O.
\end{enumerate}

The central goal of our quantum experiment is to simulate these two distinct probability distributions. By comparing the paradoxical results from the observational distribution with the true causal effect revealed by the interventional distribution, we demonstrate a resolution to Simpson's Paradox.

\subsection{Quantum Implementation of Causal Interventions}\label{ss:quantum-impl}

To translate the abstract structure of a causal DAG into a physical quantum process, we map each variable in the model to a corresponding qubit. The binary states of the variable are represented by the computational basis states of the qubit, $\ket{0}$ and $\ket{1}$. A causal arrow between two variables, for instance from a cause A to an effect B ($A \rightarrow B$), is implemented using a controlled quantum gate, where the state of the qubit representing A dictates the operation applied to the qubit representing B. In this work, we primarily use the controlled Y-rotation gate, $\text{CRY}(\theta)$, to encode these probabilistic causal links. A $\text{CRY}(\theta)$ gate applies a rotation around the Y-axis, $R_Y(\theta)$, to the target qubit (B) if and only if the control qubit (A) is in the state $\ket{1}$.

The rotation angle, $\theta$, serves as a tunable parameter that directly corresponds to the strength of the causal relationship. If the target qubit B starts in the state $\ket{0}$, the probability of it transitioning to $\ket{1}$ under the influence of the control qubit A being in state $\ket{1}$ is given by the conditional probability $P(B=1|A=1) = \sin^2(\theta/2)$. A larger angle $\theta$ thus corresponds to a stronger causal influence. For cases where the causal link is conditional on the cause being 0 (e.g., in our confounding model where males, G=0, have a higher treatment probability), the control operation is wrapped with X gates on the control qubit.

By composing these controlled-rotation gates according to a topological order of the DAG, we construct quantum circuits whose measurement statistics reproduce the SCM's joint probability distribution. To model the observational distribution, for instance, we build an \textit{observational circuit} that translates every causal arrow in the graph—including confounding paths like $G \rightarrow T$—into a corresponding gate. Crucially, as our entire construction uses only gates that are block-diagonal in the computational basis, it operates within a classical sector of the operator formalism. This ensures the circuit's semantics are mathematically identical to the classical SCM, a claim formally proven in Appendix C.

To sample from an interventional distribution, such as $P(O, G|\mathcal{DO}(T=t))$, we implement the $\mathcal{DO}$-operator by performing a procedure based on the \textit{project-prepare circuit surgery} we introduced in our prior work~\cite{kang:2025:esc-arxiv}, which for brevity we term ``circuit surgery''. Conceptually, this process involves identifying and removing all quantum gates from the observational circuit that correspond to the causal arrows pointing into the variable being intervened upon. Our implementation achieves this by directly constructing a new interventional circuit where these gates are omitted.

For an intervention on the treatment variable (T), this means removing the gates that encode the confounding path $G \rightarrow T$. After severing these incoming causal links, the state of the intervention qubit is deterministically set to the desired value; for instance, for the intervention $\mathcal{DO}(T=1)$, an X gate is applied to the treatment qubit to force its state to $\ket{1}$. The resulting interventional circuit is structurally different from the observational one and represents a new causal world where the treatment is no longer influenced by its original causes. Consequently, executing this modified circuit produces measurement outcomes that are fair samples from the post-intervention distribution $P(O, G|\mathcal{DO}(T=t))$, from which the true, unconfounded causal effect of the treatment can be calculated. The formal statement and proof sketch guaranteeing that this circuit surgery procedure is mathematically equivalent to the $\mathcal{DO}$-operation for our family of circuits is presented in Appendix D.

% ------------------------------------
\subsection{Specific Circuit Models for Simpson's Paradox}

\begin{figure*}[htbp!]
    \centering
    \begin{subfigure}[b]{0.85\textwidth}
        \includegraphics[width=\textwidth]{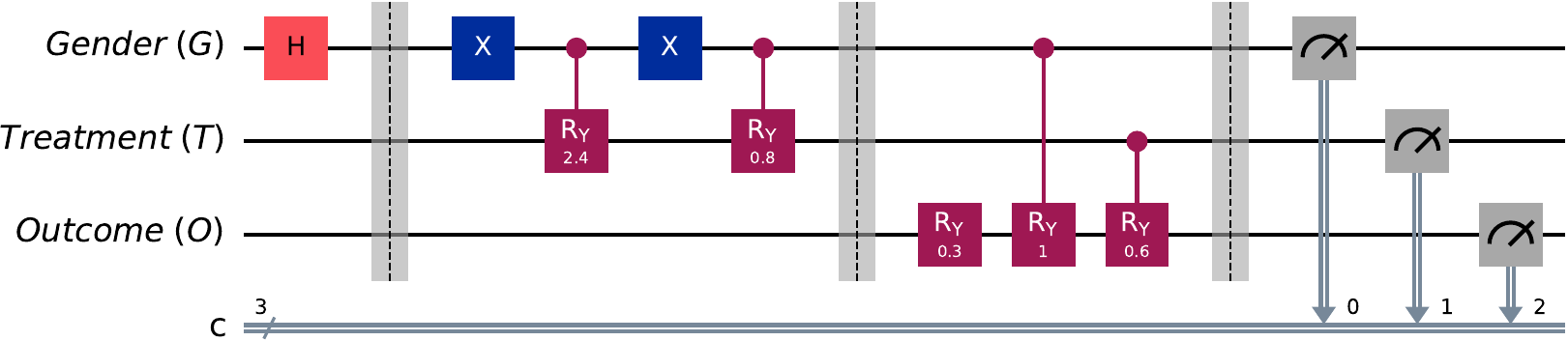}
        \caption{Observational circuit modeling the confounded system.}
        \label{fig:3q_circ_obs}
    \end{subfigure}

    \vspace{0.5cm}

    \begin{subfigure}[b]{0.7\textwidth}
        \includegraphics[width=\textwidth]{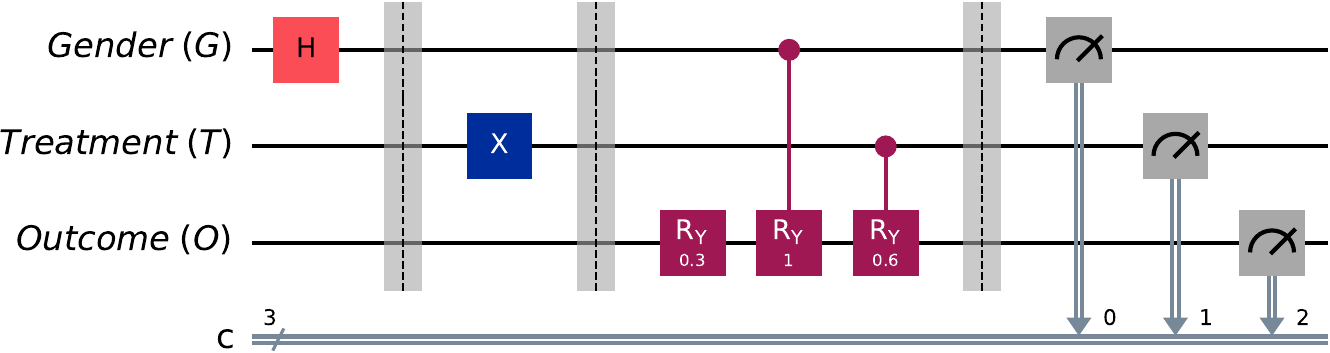}
        \caption{Interventional circuit implementing $\mathcal{DO}(T{=}1)$ via circuit surgery.}
        \label{fig:3q_circ_int}
    \end{subfigure}

    \caption{Quantum circuits for the 3-qubit Simpson model. (a) The observational circuit encodes all causal links. The dashed box highlights the gates implementing the confounding path ($G\!\to\!T$) targeted for removal via circuit surgery. (b) The resulting interventional circuit, where the confounding path has been removed and $T$ is deterministically prepared to sample from $P(O,G\mid \mathcal{DO}(T{=}1))$. All operations act within the block-diagonal (classical) sector, ensuring Born probabilities coincide with classical SCM semantics.}
    \label{fig:3q_circuits}
\end{figure*}

To validate our proposed methodology, we designed and simulated two distinct quantum circuits: a minimal 3-qubit model to clearly illustrate the core principles, and a more complex 10-qubit model to demonstrate the scalability of our approach to networks with multi-level confounding.

\subsubsection{3-Qubit Foundational Model}

This model serves as a minimal physical realization of Simpson's Paradox using a 3-qubit system. Each qubit is assigned to a core binary variable as follows: qubit $q_0$ represents Gender (G), where $\ket{0}$ corresponds to `male' and $\ket{1}$ to `female'; qubit $q_1$ represents Treatment (T), where $\ket{0}$ indicates `untreated' and $\ket{1}$ indicates `treated'; and qubit $q_2$ represents the Outcome (O), where $\ket{0}$ is a `poor' outcome and $\ket{1}$ is a `good' outcome.  The observational circuit is constructed to embody the necessary causal structure for the paradox to emerge. First, a strong confounding link ($G \rightarrow T$) is engineered such that one group (males, G=0) has a high probability ($\sim$85\%) of receiving the treatment, while the other (females, G=1) has a low probability ($\sim$15\%). Second, direct causal links are implemented for the confounder-outcome relationship ($G \rightarrow O$, where females have a better baseline outcome) and the treatment-outcome relationship ($T \rightarrow O$, where the treatment is beneficial for all subjects). The detailed structures of the resulting observational and interventional circuits for this model are depicted in Fig.~\ref{fig:3q_circuits}.  
%Conceptually, this implementation resolves the paradox by physically realizing the classical backdoor adjustment formula~\cite{pearl:2009:causality}, a correspondence that is formally stated in the Appendix (see Corollary~\ref{cor:backdoor_adjustment}).

\subsubsection{10-Qubit Scalability Model}

To demonstrate the utility of our method on more complex, realistic problems, we developed a 10-qubit model simulating a multi-level healthcare scenario. The model features a main causal chain where demographic variables influence treatment decisions, which in turn affect a cascade of healthcare quality factors and the final patient outcome (e.g., Age $\rightarrow$ Income $\rightarrow \dots \rightarrow$ Treatment $\rightarrow \dots \rightarrow$ Outcome). The circuit also includes additional confounding paths (e.g., Age $\rightarrow$ Insurance) to create a deeply confounded system where the true causal effect of the treatment is obscured by multiple interacting variables. This model serves as a more challenging testbed to assess the ability of our quantum causal inference framework to disentangle complex causal relationships that are difficult to address with simple stratification.

\subsubsection{Circuit-Based Resolution of Simpson's Paradox}
To resolve Simpson's Paradox, one must compute the true causal effect using the backdoor adjustment formula~\cite{pearl:2009:causality}, which isolates the effect of $X$ on $Y$ by accounting for the confounder $Z$:
\[
P(y\mid \DO(x))=\sum_{z} P(y\mid x,z)\,P(z).
\]
Our quantum framework is designed to be a direct physical realization of this adjustment. The observational circuit first provides the necessary population statistics, such as the distribution $P(z)$. Then, the interventional circuit, created via circuit surgery on $X$, produces samples from the post-interventional distributions, which allows for the calculation of the conditional probabilities $P(y \mid x, z)$. 

By combining these results according to the formula, our method is mathematically guaranteed to yield the true causal effect, $P(y \mid \DO(x))$, thereby resolving the paradox. A formal statement and proof that our compiled circuit primitives correctly implement this compositional causal query is provided in Appendix E.

%%%%%%%%%%%%%%%%%%%%%%%%%%%%%%%%%%%%%%%%%%%%%%%%%%%%%%%%%%
\section{Experiments and Results}\label{s:results}

In this section we report the behavior of our quantum causal intervention framework on both the 3-qubit Simpson-type model and the 10-qubit healthcare network, combining noiseless simulations with execution on the IonQ Aria NISQ processor.

\subsection{Experimental Setup}

All experiments were designed and executed within a Python 3 environment, utilizing the Qiskit open-source framework (v1.4.3)~\cite{javadiabhari:2024:qiskit} for quantum circuit construction and execution. Data analysis and visualization were performed using the standard scientific Python libraries, including Numpy~\cite{harris:2020:numpy}, Scipy~\cite{virtanen:2020:scipy}, and Matplotlib~\cite{hunter:2007:matplotlib}. The data that support the findings of this study, as well as the source code for the experiments and analysis, are publicly available in a GitHub repository at https://github.com/pilsungk/quantum-simpson.

Our study involved two types of backends: a high-performance classical simulator and a physical quantum processing unit (QPU). All simulations were performed using the \texttt{AerSimulator} from the \texttt{qiskit\_aer} package. The 3-qubit foundational model simulation was conducted over 30 trials with 15,000 shots per circuit to achieve high statistical precision. The 10-qubit scalability model was simulated for 10 trials with 15,000 shots each. The experimental validation was performed on the IonQ Aria trapped-ion quantum computer, accessed via the \texttt{qiskit-ionq} provider. The hardware experiment consisted of 3 successful trials with 1024 shots per circuit.

To ensure statistical significance, all reported 95\% confidence intervals for the treatment effects are derived from the distribution of the outcomes across these independent trials. The intervals are calculated using a normal approximation, defined as the sample mean $\pm 1.96$ times the standard error of the mean.

\subsection{Result 1: Simpson's Paradox on a 3-Qubit Model: Simulation and Experimental Validation}

\begin{figure}
    \centering
    \includegraphics[width=0.48\textwidth]{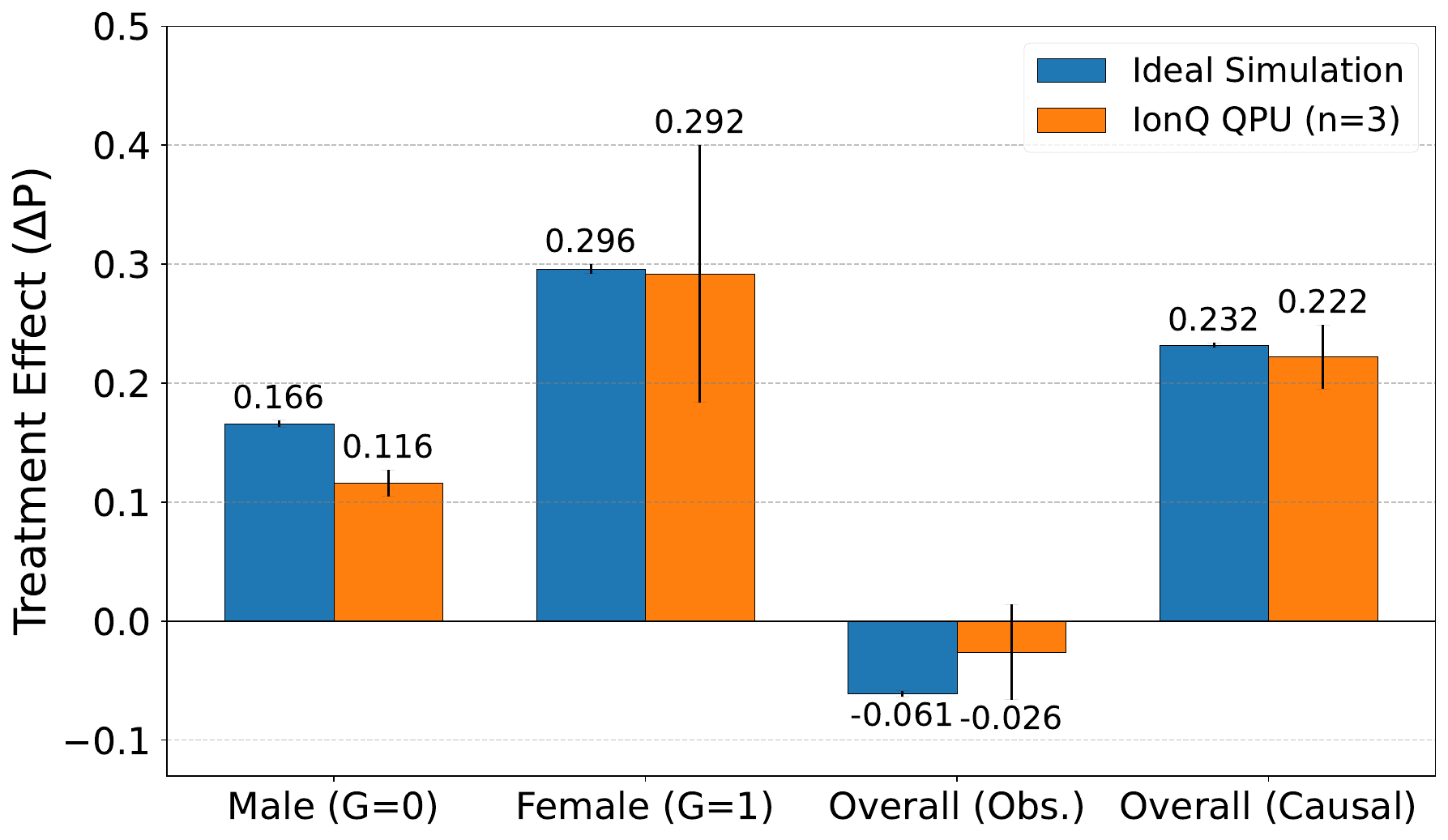}
    \caption{Comparison of 3-qubit Simpson's Paradox results from ideal simulation (blue bars) and the IonQ Aria QPU (orange bars). Both the simulation and the QPU experiment demonstrate the core phenomenon: the treatment effect is positive within the male and female subgroups, but reverses to become negative when aggregated (`Overall (Obs.)'). In both cases, the quantum causal intervention ($\mathcal{DO}$-operation) successfully resolves the paradox by revealing a robustly positive true causal effect (`Overall (Causal)'). Error bars represent 95\% confidence intervals, calculated over 30 trials for the simulation and 3 successful trials for the QPU experiment. The QPU results qualitatively reproduce the simulation's trend, validating the method on real hardware, while the larger error bars on the QPU data reflect the impact of physical noise. }
    \label{fig:3q_barchart_compact}
\end{figure}

We first demonstrate our framework on the 3-qubit foundational model, performing both an ideal noise-free simulation and an experimental validation on a quantum processor. This dual approach allows us to first establish the theoretical correctness of our method in a baseline scenario and then to directly assess its practical performance and the impact of real-world hardware noise. The complete results, which directly compare the simulation and the IonQ QPU execution, are presented in Table~\ref{tab:3q_sim_vs_qpu_compact} and Fig.~\ref{fig:3q_barchart_compact}. As we will show, the experimental results successfully reproduce the key features of the ideal simulation, confirming the viability of our approach.

\begin{table*}[htbp]
\centering
\caption{Simpson-type treatment effects in the 3-qubit model from an ideal simulator and the IonQ Aria QPU (3 trials, 1024 shots per circuit). In both backends the treatment effect is positive within each gender subgroup but negative in the aggregated observational analysis, while the quantum $\mathcal{DO}$-intervention recovers a positive causal effect with wider confidence intervals on the QPU reflecting hardware noise.}
%\caption{Comparison of 3-qubit Simpson's Paradox results from the ideal simulator and the IonQ Aria QPU (n=3 trials, 1024 shots per circuit). In both the simulation and the QPU experiment, the treatment effect is positive within the male and female subgroups, but the aggregated observational effect becomes negative, demonstrating the paradox. The quantum causal intervention ($\mathcal{DO}$-operation) successfully resolves this paradox in both cases by revealing the true, positive causal effect. The QPU data qualitatively reproduces the simulation results, while its wider confidence intervals highlight the impact of real-world hardware noise on the statistical outcomes.  }
\label{tab:3q_sim_vs_qpu_compact}
\begin{tabular*}{\textwidth}{l @{\extracolsep{\fill}} cccc} 
\toprule
\multirow{2}{*}{\textbf{Analysis Group}} & \multicolumn{2}{c}{\textbf{Ideal Simulation}} & \multicolumn{2}{c}{\textbf{IonQ QPU}} \\
\cmidrule(lr){2-3} \cmidrule(lr){4-5}
 & \textbf{Effect Size ($\Delta P$)} & \textbf{95\% CI} & \textbf{Effect Size ($\Delta P$)} & \textbf{95\% CI} \\
\midrule
Observational, Male (G=0)   & +0.166 & [+0.163, +0.169] & +0.116 & [+0.105, +0.127] \\
Observational, Female (G=1) & +0.296 & [+0.292, +0.300] & +0.292 & [+0.184, +0.400] \\
\midrule
Observational, Overall      & \textbf{-0.061} & \textbf{[-0.063, -0.058]} & \textbf{-0.026} & \textbf{[-0.066, +0.014]} \\
Causal, Overall ($\mathcal{DO}$)   & +0.232 & [+0.230, +0.234] & +0.222 & [+0.195, +0.249] \\
\bottomrule
\end{tabular*}
\end{table*}

%% \begin{table}[htbp]
%% \centering
%% \caption{Comparison of observational and causal effects of the treatment in the 3-qubit simulation. While the treatment effect is positive within both subgroups (Male and Female), the aggregated overall effect is negative, demonstrating Simpson's Paradox. The causal intervention ($\mathcal{DO}$-operation) resolves this by revealing the true positive effect.}
%% \label{tab:3q_results}
%% \begin{tabular}{l c c c}
%% \toprule
%% \textbf{Group} & \textbf{Analysis Type} & \textbf{Effect Size ($\Delta P$)} & \textbf{95\% CI} \\
%% \midrule
%% Male ($G=0$)   & Observational & +0.166 & [+0.163, +0.169] \\
%% Female ($G=1$) & Observational & +0.296 & [+0.292, +0.300] \\
%% \midrule
%% Overall      & \textbf{Observational} & \textbf{-0.061} & \textbf{[-0.063, -0.058]} \\
%% Overall      & \textbf{Causal ($\mathcal{DO}$)}    & \textbf{+0.232} & \textbf{[+0.230, +0.234]} \\
%% \bottomrule
%% \end{tabular}
%% \end{table}

In the simulation, the observational analysis reveals positive treatment effects for both male ($+0.166$) and female ($+0.296$) subgroups, which reverse to a negative overall effect of $-0.061$. Critically, this paradoxical structure is qualitatively reproduced on the quantum processor, which measured subgroup effects of $+0.116$ and $+0.292$ respectively, against a negative-trending overall effect of $-0.026$. Furthermore, the causal resolution is robustly demonstrated in both environments. The quantum causal intervention yields a true average causal effect (ACE) of $+0.232$ in the simulation and a consistent $+0.222$ on the QPU. A direct comparison highlights the impact of hardware noise; while the strong causal signal remains statistically significant in the QPU data (95\% CI $[0.195, 0.249]$), the confidence interval for the weaker observational overall effect, $[-0.066, +0.014]$, now includes zero, underscoring the challenge of observing subtle statistical phenomena on noisy hardware.

\subsection{Result 2: Scalability on a Complex Network (Simulation)}

To assess the scalability and practical utility of our framework beyond minimal models, we applied it to a complex 10-qubit causal network simulating a multi-level healthcare scenario. As described in the Methodology section, this model was designed as a more challenging testbed, featuring a long causal chain and multiple confounding pathways that are difficult to disentangle using conventional statistical methods. The simulation results, summarized in Table~\ref{tab:10q_results}, demonstrate that while the observational data suggests a positive treatment effect, it significantly underestimates the true causal impact---a bias that our interventional method successfully quantifies.

\begin{table}[htbp]
\centering
\caption{Comparison of treatment effects under various analysis levels in the 10-qubit system. The 'Bias' column represents the deviation of each calculated effect from the true causal effect (+0.486) discovered via $\mathcal{DO}$-calculus.}
\label{tab:10q_results}
\begin{tabular}{l c c c}
\toprule
\textbf{Analysis Level} & \textbf{Effect Size ($\Delta P$)} & \textbf{95\% CI} & \textbf{Bias} \\
\midrule
Overall Observational    & +0.377 & [0.371, 0.383] & -0.109 \\
\addlinespace 
Stratified by Age        & +0.497 & [0.491, 0.503] & +0.011 \\
Stratified by Region       & +0.406 & [0.396, 0.416] & -0.080 \\
\midrule
\textbf{Causal Intervention} & \textbf{+0.486} & \textbf{[0.481, 0.490]} & \textbf{0.000} \\
\bottomrule
\end{tabular}
\end{table}

\begin{figure}
    \centering
    \includegraphics[width=0.48\textwidth]{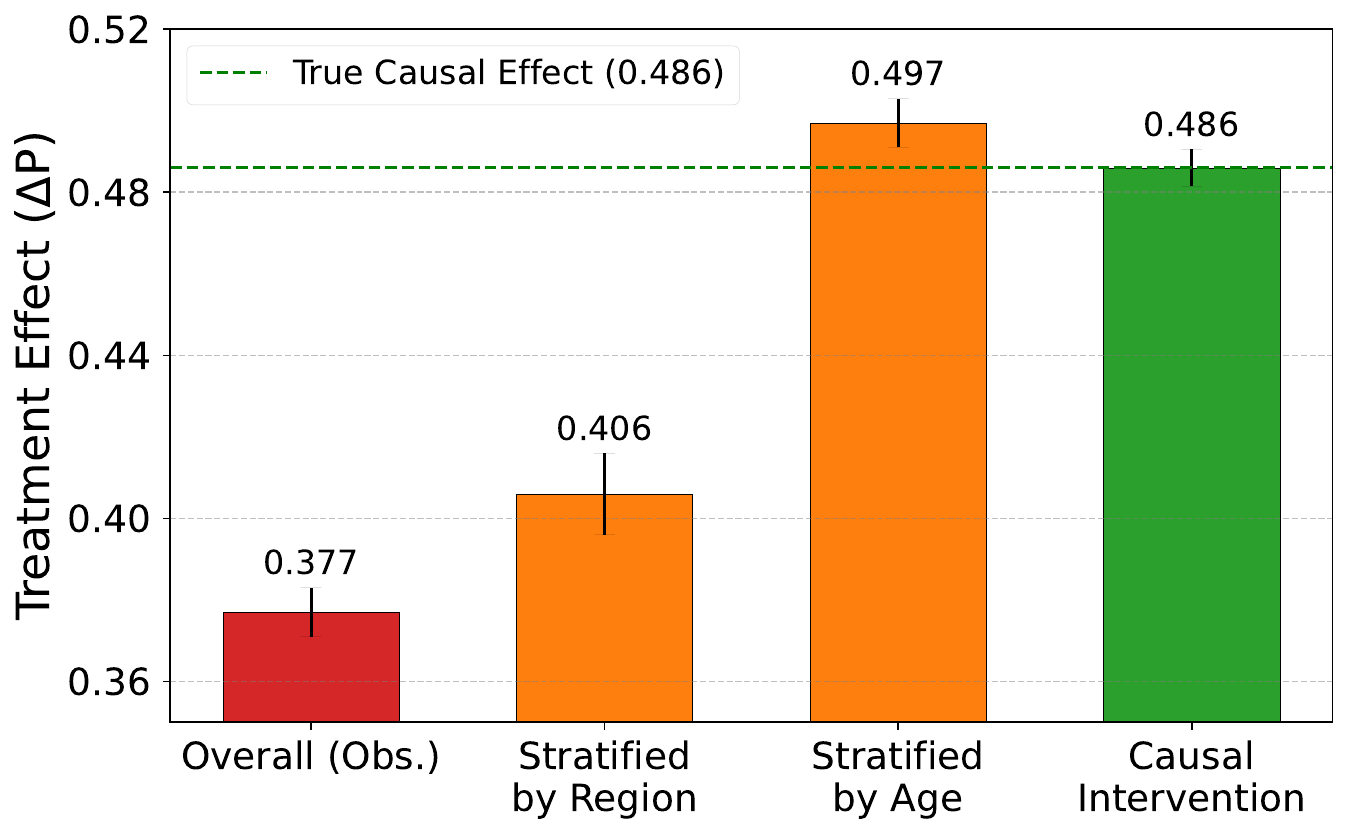}
    \caption{Quantifying confounding bias in the 10-qubit healthcare simulation. This figure compares the treatment effect ($\Delta P$) calculated from raw observational data against results from conventional stratification and our quantum causal intervention. The observational data (`Overall (Obs.)') significantly underestimates the true causal effect, yielding an effect of +0.377, whereas the true effect revealed by the intervention (`Causal Intervention') is +0.486. Simple stratification by a single variable provides only a partial and inconsistent correction; stratifying by Region yields an effect of +0.406, while stratifying by the stronger confounder, Age, yields +0.497. %This demonstrates the robustness of the quantum $\mathcal{DO}$-calculus approach in isolating the true causal effect from complex confounding structures. The horizontal dashed line indicates the true causal effect for reference. Error bars represent 95\% confidence intervals calculated over 10 trials.
    }
    \label{fig:10q_enhanced}
\end{figure}

The detailed results, presented in Table~\ref{tab:10q_results} and visualized in Fig.~\ref{fig:10q_enhanced}, reveal the insufficiency of conventional statistical controls in this complex network. The overall observational data yields a treatment effect of $+0.377$ (95\% CI $[0.371, 0.383]$). Attempting to correct for this by stratifying on a single confounding variable provides only a partial and inconsistent solution; stratification by the `Region' variable adjusts the effect to $+0.406$, while stratification by the more influential `Age' variable adjusts it to $+0.497$. In contrast, our quantum causal intervention method, which accounts for all confounding paths simultaneously, isolates the true average causal effect (ACE) to be $+0.486$ (95\% CI $[0.481, 0.490]$). This demonstrates that in systems with multi-level confounding, where the choice of a single stratification variable can lead to different conclusions, our quantum $\mathcal{DO}$-calculus implementation provides a robust and definitive means of identifying the true causal impact.

For full reproducibility, the specific gate parameters (rotation angles $\theta$) used to construct both the 3-qubit and 10-qubit causal models, along with the compilation recipe used to derive them from conditional probabilities, are detailed in Appendix A and F.

%%%%%%%%%%%%%%%%%%%%%%%%%%%%%%%%%%%%%%%%%
\section{Discussions}\label{s:discuss}

In this work, we have demonstrated a quantum algorithmic framework for causal inference, validating its ability to reproduce the classical resolution of Simpson-type paradoxes and quantify confounding bias through both high-fidelity simulations and a small-scale experimental demonstration on a quantum processor. Beyond the specific numerical outcomes, our findings support a deeper interpretation of quantum circuits as direct physical analogues for SCMs. As established in our foundational work~\cite{kang:2025:esc-arxiv}, this framework uses superposition to represent the probabilistic nature of causal variables. Crucially, both the arrows representing true causal effects and those forming the confounding structures that induce spurious correlations are realized by the same physical mechanism: entanglement generated via controlled gates. From this perspective, the $\mathcal{DO}$-operator is implemented via the circuit surgery procedure---a structural modification that creates a new physical system corresponding to the interventional world. Building on this interpretation, the following sections will analyze the specific biases observed in our models, position our work relative to classical methods and its implications for AI, and discuss the limitations and future directions of this approach.

\subsection{Interpretation of Confounding Bias in Causal Models}

Our results provide a tale of two confounding scenarios, illustrating the different ways in which observational data can be misleading. In the 3-qubit foundational model, we observed a classic case of Simpson's Paradox where the confounding was strong enough to completely reverse the sign of the treatment effect. The data showed a statistically significant positive effect within subgroups, which flipped to a negative effect in the aggregated observational data (e.g., effect = $-0.061$ in simulation). Our quantum intervention correctly identified this reversal and recovered the true, positive causal effect.

The 10-qubit model presented a more subtle, and perhaps more realistic, scenario. Here, the confounding was not strong enough to reverse the sign of the effect, but it nonetheless created a significant quantitative bias. The observational data suggested a positive treatment effect of $+0.377$, which is a substantial underestimation of the true causal effect of $+0.486$ revealed by our quantum intervention. The practical implications of this latter finding are profound. A model or policy based purely on the observational data would have undervalued the treatment's true efficacy by more than 22\% ($0.109 / 0.486$). In a real-world context, such as evaluating a new medical intervention or a public policy, an error of this magnitude could lead to the premature rejection of a beneficial treatment or the misallocation of critical resources. This highlights the critical need for computational tools that can move beyond simple correlations to robustly estimate true causal impacts, a role our quantum framework is designed to fill.

\subsection{Generalization to a Compositional Causal Calculus}

The results presented for Simpson's paradox are a specific instance of a more general capability of our framework to act as a compiler for causal reasoning. As detailed in Sec.~\ref{ss:quantum-impl}, each primitive operation of the $\mathcal{DO}$-calculus is assigned a fixed circuit-level semantics: $\mathcal{DO}$-interventions are implemented via circuit surgery on the observational circuit, conditioning on $Z=z$ corresponds to measuring the $Z$-register and classically post-selecting the desired outcome, and marginalization over $Z$ is realized by summing the resulting measurement statistics. 

The key observation is that these primitives compose. Complex causal queries such as the backdoor adjustment formula
\[
P(y \mid \mathcal{DO}(x)) \;=\; \sum_{z} P(y \mid x,z)\,P(z)
\]
used in our Simpson-type example are built from exactly these ingredients. As formally shown in Appendix~E, any valid expression obtained from an SCM by combining interventions, conditioning, and marginalization in the sense of the $\mathcal{DO}$-calculus is faithfully reproduced by the corresponding composition of our circuit primitives. This positions our framework not merely as a tool for resolving a single paradox, but as an experimentally validated mapping from abstract causal calculus to executable quantum circuits.

% \subsection{Generalization to a Compositional Causal Calculus}
% 
% The results presented for Simpson's Paradox are a specific instance of a more general capability of our framework to act as a compiler for causal reasoning. This approach translates high-level causal queries into executable quantum circuits by providing circuit-level semantics for the three fundamental primitives of the causal calculus. First, an intervention ($\mathcal{DO}$) is compiled into our circuit surgery procedure, where confounding pathways are physically removed from the circuit's structure. Second, a conditioning operation, such as on a variable $Z=z$, is compiled into a measurement on the corresponding qubit followed by classical post-selection of the desired outcomes. Third, marginalization, such as summing over $Z$, is compiled into the classical post-processing of summing the counts across all states of the qubit for $Z$.
% 
% The power of this approach lies in its compositionality. Complex causal queries, such as the backdoor adjustment formula ($P(y|\mathcal{DO}(x)) = \sum_z P(y|x,z)P(z)$) used to resolve Simpson's Paradox, are composed of these exact primitives. As formally shown in Appendix E, any valid combination of these operations in a classical SCM is faithfully reproduced by the corresponding composition of our circuit primitives. This establishes our framework not merely as a tool for resolving a specific paradox, but as a robust and verifiable compiler that translates the abstract language of causal calculus into concrete, physical operations on a quantum processor.
% 
\subsection{Implications and Comparison with Classical Methods}

The ability to perform physical causal interventions on a quantum computational substrate has significant implications across several scientific and technological domains. In machine intelligence, our framework offers a new tool for enhancing explainability and algorithmic fairness. By moving beyond correlational models, it provides a means to probe the true causal drivers behind a model's decisions, helping to identify and mitigate biases learned from confounded data. Beyond AI, this approach could be applied to complex network problems in fields such as systems biology, materials science, and econometrics, where simulating interventions to understand the effects of a specific gene, material property, or policy is a central challenge.

This computational approach to intervention stands in contrast to the gold standard of classical experimental design: the RCT. While RCTs achieve causal identification by physically randomizing subjects, they can be prohibitively expensive, time-consuming, unethical (e.g., forcing subjects to adopt harmful behaviors), or physically impossible (e.g., re-running historical events). Our quantum framework, acting as a ``computational laboratory,'' allows such interventions to be simulated on a validated causal model. This enables researchers to explore a vast range of ``what-if'' scenarios that are inaccessible to traditional experimental methods.

It is important to position our quantum approach relative to established classical methods for causal inference, such as covariate adjustment based on the back-door criterion or instrumental variable analysis~\cite{pearl:1995:do}. For systems where the causal graph is known and the number of variables is small, these statistical methods are both efficient and sufficient. The motivation for a quantum approach, however, lies in scenarios where the system's complexity poses a significant challenge to classical methods. For causal networks with a large number of variables and highly intricate dependencies, the state space can grow exponentially, which can make it computationally expensive to represent and sample from the full joint probability distribution classically. A quantum approach, by leveraging properties like superposition and entanglement, offers a fundamentally different paradigm for representing and manipulating these high-dimensional causal structures. In this work we remain agnostic about any potential speedups and focus instead on showing that SCM-style graph surgery and $\mathcal{DO}$-operations can be faithfully implemented and empirically validated on present-day NISQ processors.

%While the current work does not demonstrate a quantum advantage, our successful experimental validation serves as a foundational proof-of-principle, establishing a pathway toward a potential ``quantum advantage for causal inference'' on future, fault-tolerant quantum hardware.

\subsection{Limitations and Future Directions}

While our work establishes a viable framework for quantum causal inference, it is important to clarify its scope and current limitations. The experiments presented in this paper focus on causal effect estimation and therefore presuppose that the underlying causal graph is known a priori, which is a standard assumption for applying the $\mathcal{DO}$-calculus. However, this does not mean the framework is inapplicable when the true graph is unknown. Indeed, the ability of our circuit-based approach to act as a computational laboratory allows it to serve as a powerful tool for causal model testing. Researchers can formulate several competing causal hypotheses, compile each into a distinct quantum circuit, and compare the simulated observational data from each circuit against real-world data to determine the most plausible causal model. While a full exploration of such a quantum-enhanced hypothesis testing workflow is beyond the scope of this paper, it represents a significant direction for future research.

Second, while we have provided a successful proof-of-principle on current quantum hardware, a significant gap remains between the performance on small-scale problems and the requirements for achieving a practical quantum advantage on large-scale problems with NISQ devices.

Finally, throughout this work we deliberately restrict attention to block-diagonal circuits that act within a classical sector of the Hilbert space; relaxing this constraint to allow non-block-diagonal dynamics with genuinely quantum confounders is a natural direction for future work and would clarify in which regimes quantum resources can offer benefits beyond classically simulable SCMs.

Taken together, these limitations highlight several concrete avenues for future research. A primary direction is the development of quantum algorithms for causal discovery, which would aim to learn the causal graph itself from quantum observational and interventional data, thereby removing our framework's main prerequisite. Another critical research path is the design of noise-robust causal inference circuits. This could involve co-designing algorithms with specific hardware properties in mind or integrating quantum error mitigation and correction techniques to improve the fidelity of results on current and future NISQ processors. Progress in these areas will be crucial for moving quantum causal inference from a small-scale experimental validation, as demonstrated here, to a powerful tool for scientific discovery in complex systems.

%%%%%%%%%%%%%%%%%%%%%%%%%%%%%%%%%%%%%%%%%%%%%%%%
\section{Conclusion}\label{s:conc}

In this work, we addressed the fundamental challenge of causal inference in machine intelligence by applying and experimentally validating a quantum algorithmic framework for implementing Pearl's $\mathcal{DO}$-calculus. By representing causal networks as quantum circuits and performing interventions via circuit surgery, our method provides a direct, physically-grounded means of distinguishing true causal effects from spurious correlations. Our framework successfully resolved Simpson's Paradox in a 3-qubit model, demonstrated its potential to quantify bias in a complex 10-qubit simulation, and, most critically, was validated in a proof-of-principle experiment on an IonQ quantum computer. This work establishes a practical pathway for quantum causal inference, offering a new class of tools for building more robust, fair, and explainable AI systems. By bridging the mathematical rigor of causal models with the physical dynamics of quantum systems, our approach opens a new frontier for research at the intersection of quantum computing and machine intelligence.

\section*{Acknowledgments}
This work was supported by the National Research Foundation of Korea (NRF) grant funded by the Korea government (Ministry of Science and ICT (MSIT)), grant number 2020R1F1A1067619. This research was supported by `Quantum Information Science R\&D Ecosystem Creation' through NRF funded by the Korea government (MSIT), grant number 2020M3H3A1110365. The authors acknowledge Google's Gemini for its assistance during the preparation of this manuscript. The AI model was utilized for refining the language and structure of several paragraphs, and improving Python code for circuit visualization.


\appendices 
\section{Quantum Gate Parameters for the Causal Models}
\label{app:gate_params}

The specific rotation angles ($\theta$) for all gates used to construct the 3-qubit and 10-qubit causal models are detailed in Table~\ref{tab:gate_parameters}. 

\begin{table*}[htbp!]
\centering
\caption{Quantum gate parameters for the causal models. All rotation angles $\theta$ are given in radians.}
\label{tab:gate_parameters}
\begin{tabular*}{0.9\textwidth}{l @{\extracolsep{\fill}} l l c} 
\toprule
\textbf{Causal Link / Base Rate} & \textbf{Condition} & \textbf{Quantum Gate} & \textbf{Angle ($\theta$)} \\
\midrule
\multicolumn{4}{c}{\textit{3-Qubit Foundational Model}} \\
\midrule
Gender (G) Distribution      & -                  & $H(q_0)$              & -                \\
Gender (G) $\rightarrow$ Treatment (T) & Male (G=0)         & $\text{CRY}(\theta)$   & 2.4              \\
                             & Female (G=1)       & $\text{CRY}(\theta)$   & 0.8              \\
Outcome (O) Base Rate        & -                  & $R_Y(\theta)$         & 0.3              \\
Gender (G) $\rightarrow$ Outcome (O)   & Female (G=1)       & $\text{CRY}(\theta)$   & 1.0              \\
Treatment (T) $\rightarrow$ Outcome (O) & -                  & $\text{CRY}(\theta)$   & 0.6              \\
\midrule
\multicolumn{4}{c}{\textit{10-Qubit Scalability Model}} \\
\midrule
Age (A) Base Rate                        & -                  & $R_Y(\theta)$         & 1.0              \\
Age (A) $\rightarrow$ Income (I)         & Old (A=1)          & $\text{CRY}(\theta)$   & 0.8              \\
Income (I) $\rightarrow$ Region (R)      & High Income (I=1)  & $\text{CRY}(\theta)$   & 1.2              \\
Region (R) $\rightarrow$ Gender Bias (G) & Rural (R=0)        & $\text{CRY}(\theta)$   & 1.0              \\
Treatment (T) Base Rate                  & -                  & $R_Y(\theta)$         & 0.2              \\
Age (A) $\rightarrow$ Treatment (T)      & Old (A=1)          & $\text{CRY}(\theta)$   & 1.2              \\
Income (I) $\rightarrow$ Treatment (T)   & High Income (I=1)  & $\text{CRY}(\theta)$   & 1.0              \\
Gender Bias (G) $\rightarrow$ Treatment (T) & High Bias (G=1)    & $\text{CRY}(\theta)$ (Negative effect) & 1.4 \\
Insurance (S) Base Rate                  & -                  & $R_Y(\theta)$         & 0.3              \\
Treatment (T) $\rightarrow$ Insurance (S) & Treated (T=1)      & $\text{CRY}(\theta)$   & 0.8              \\
Age (A) $\rightarrow$ Insurance (S)      & Old (A=1)          & $\text{CRY}(\theta)$   & 0.4              \\
Hospital (H) Base Rate                   & -                  & $R_Y(\theta)$         & 0.4              \\
Insurance (S) $\rightarrow$ Hospital (H) & Premium (S=1)      & $\text{CRY}(\theta)$   & 0.6              \\
Region (R) $\rightarrow$ Hospital (H)    & Urban (R=1)        & $\text{CRY}(\theta)$   & 0.5              \\
Doctor (D) Base Rate                     & -                  & $R_Y(\theta)$         & 0.5              \\
Hospital (H) $\rightarrow$ Doctor (D)    & Advanced (H=1)     & $\text{CRY}(\theta)$   & 0.4              \\
Gender Bias (G) $\rightarrow$ Doctor (D) & High Bias (G=1)    & $\text{CRY}(\theta)$   & 0.3              \\
Outcome (O) Base Rate                    & -                  & $R_Y(\theta)$         & 0.1              \\
Age (A) $\rightarrow$ Outcome (O)        & Young (A=0)        & $\text{CRY}(\theta)$   & 0.8              \\
Region (R) $\rightarrow$ Outcome (O)     & Rural (R=0)        & $\text{CRY}(\theta)$   & 0.6              \\
Treatment (T) $\rightarrow$ Outcome (O)  & Treated (T=1)      & $\text{CRY}(\theta)$   & 1.2              \\
Doctor (D) $\rightarrow$ Outcome (O)     & Senior (D=1)       & $\text{CRY}(\theta)$   & 0.5              \\
Hospital (H) $\rightarrow$ Outcome (O)   & Advanced (H=1)     & $\text{CRY}(\theta)$   & 0.4              \\
Satisfaction (F) Base Rate               & -                  & $R_Y(\theta)$         & 0.3              \\
Outcome (O) $\rightarrow$ Satisfaction (F) & Good (O=1)         & $\text{CRY}(\theta)$   & 0.8              \\
Doctor (D) $\rightarrow$ Satisfaction (F)  & Senior (D=1)       & $\text{CRY}(\theta)$   & 0.3              \\
\bottomrule
\end{tabular*} 
\end{table*}

%%%%%%%%%%%%%%%%%%%%%%%%%%%%%%%%%%%%%%%%%%%%%%%%%%%%%
\section{Theoretical Framework and Notation}
\label{app:framework_and_notation}

In this appendix, we provide the formal definitions and theoretical guarantees that underpin the quantum causal inference framework used in the main text. We begin by establishing the core notation.

\noindent\textbf{Setting and Notation.} Let $\mathcal{M}=(G,\{P(X_i \mid \mathrm{Pa}(X_i))\}_{i=1}^n)$ be a binary structural causal model (SCM) with a DAG $G$ over variables $X_1,\dots,X_n$ and parents $\mathrm{Pa}(X_i)$. 
We compile $\mathcal{M}$ into a quantum circuit $\mathcal{C}(G,\Theta)$ on $n$ qubits by assigning one qubit per variable and, for each $i$ in a topological order of $G$, appending a (multi-)controlled rotation that realizes the local conditional:
\begin{align*}
    \sin^2\!\big(\tfrac{\theta_i(pa)}{2}\big) &= P(X_i{=}1 \mid \mathrm{Pa}(X_i){=}pa), \\
    U_i(pa) &:= \mathrm{CRY}_{i}\!\big(\theta_i(pa)\big),
\end{align*}
where the controls are the qubits corresponding to $\mathrm{Pa}(X_i)$ configured at pattern $pa$ and the target is qubit $i$. 
Let $\Theta = \{\theta_i(pa)\}$ be the set of all such rotation angle parameters.
Let $U := \prod_i U_i$ (applied in a topological order), with input state $\ket{0}^{\otimes n}$ and computational-basis measurement on all qubits.

%%%%%%%%%%%%%%%%%%%%%%%%%%%%%%%%%%%%%%%%%%%%%%%%%
\section{Operator–Theoretic Embedding (Classical Diagonal Sector)}\label{app:embedding}

\noindent\textbf{Definition (Embedding).}
For a finite binary SCM on DAG \(G\), assign one qubit per variable with computational basis \(\{\lvert 0\rangle,\lvert 1\rangle\}\).
For each arrow \(A\!\to\!B\), implement a controlled rotation on \(B\) that is active only on a specified control pattern of \(A\) (and other parents if present).
Readout uses computational–basis projectors.

\medskip
\noindent\textbf{Lemma (Diagonal/classical form).}\label{lem:diag}
Under this embedding, all instruments are block–diagonal in the computational basis. Consequently, the global state remains a classical mixture over basis states and
\(\operatorname{Born}(\rho,E)=\Tr(\rho E)\) reduces to classical sums; i.e., Born probabilities coincide with classical SCM marginals.\\

\noindent\emph{Proof sketch.} Controlled rotations conditioned on disjoint classical patterns act on orthogonal subspaces; no off–diagonal coherences are created across patterns. Composition preserves block–diagonality, hence Born probabilities equal classical marginals. \(\square\)

%%%%%%%%%%%%%%%%%%%%%%%%%%%%%%%%%%%%%%%%%%%%%%%%%
\section{Correctness of Circuit Surgery: Proposition and Proof Sketch}\label{app:surgery}

\begin{proposition}[Correctness of circuit surgery for the compiled family]
\label{prop:obs-do-correctness}
For the circuit $\mathcal{C}(G,\Theta)$ constructed above, the following hold.
\begin{enumerate}
\item \textbf{Observational correctness.} The measurement distribution of $U\ket{0}^{\otimes n}$ in the computational basis equals the SCM's observational joint:
\[
\Pr_{\mathcal{C}}\big[X{=}x\big]
=\prod_{i=1}^n P\big(x_i \mid x_{\mathrm{Pa}(X_i)}\big)
=: P_{\mathcal{M}}(x).
\]
\item \textbf{Interventional correctness via circuit surgery.} 
Let $S\subseteq\{1,\dots,n\}$ and $s\in\{0,1\}^{|S|}$. 
Define the \emph{circuit-surgery} intervention $\mathcal{C}^{\mathcal{DO}(S{=}s)}$ by (i) deleting all controls into targets in $S$ (i.e., removing gates that encode $P(X_j\mid \mathrm{Pa}(X_j))$ for $j\in S$), and (ii) fixing qubits in $S$ to the computational basis state $\ket{s}$ (e.g., by applying $X$ on those with $s_j{=}1$), while leaving all downstream gates unchanged. 
Then the measurement distribution of $\mathcal{C}^{\mathcal{DO}(S{=}s)}$ equals the truncated-factorization interventional distribution of the SCM:
\begin{align*}
    \Pr_{\mathcal{C}^{\mathcal{DO}}}\big[X{=}x\big]
    &= \mathbf{1}[x_S{=}s]\prod_{i\notin S}P\big(x_i \mid x_{\mathrm{Pa}(X_i)\setminus S},\, s\big) \\
    &=: P_{\mathcal{M}}\big(x \,\big|\, \mathcal{DO}(S{=}s)\big).
\end{align*}
\end{enumerate}
\end{proposition}

\begin{proof}[Proof sketch]
We prove (1) and (2) by induction over a topological order of $G$.

\emph{(1) Observational case.}
For the base variable $X_{i_1}$ (with no parents), $U_{i_1}=\mathrm{RY}(\theta_{i_1})$ prepares 
$\sqrt{1-p}\ket{0}+\sqrt{p}\ket{1}$ with $p=P(X_{i_1}{=}1)$, so Born's rule yields the correct marginal.
Assume after applying $U_{i_1},\dots,U_{i_k}$ the joint marginal on $\{X_{i_1},\dots,X_{i_k}\}$ equals $\prod_{\ell=1}^k P(x_{i_\ell}\mid x_{\mathrm{Pa}(X_{i_\ell})})$.
For the next node $X_{i_{k+1}}$, the controlled $\mathrm{RY}$ acts \emph{conditionally} on the already prepared parents, rotating the target amplitude so that 
$\sin^2(\theta_{i_{k+1}}(pa)/2)=P(X_{i_{k+1}}{=}1\mid pa)$ for each computational branch $pa=x_{\mathrm{Pa}(X_{i_{k+1}})}$.
Linearity ensures that the post-gate amplitude-squared on each branch multiplies the previous branch probability by $P(x_{i_{k+1}}\mid pa)$.
Thus the joint after $U_{i_{k+1}}$ factors correctly, completing the induction.

\emph{(2) Interventional (circuit-surgery) case.}
Deleting all incoming controls to $S$ and fixing $X_S{=}s$ precisely implements the \emph{truncated factorization}:
the factors $P(X_j\mid \mathrm{Pa}(X_j))$ for $j\in S$ are replaced by point masses $\mathbf{1}[x_j{=}s_j]$, and all descendants retain their conditionals with $S$ clamped to $s$.
By the same inductive argument as above, each downstream controlled rotation now conditions on $(x_{\mathrm{Pa}(X_i)\setminus S}, s)$, yielding the product
$\mathbf{1}[x_S{=}s]\prod_{i\notin S}P(x_i\mid x_{\mathrm{Pa}(X_i)\setminus S}, s)$, which equals $P_{\mathcal{M}}(x\mid \mathcal{DO}(S{=}s))$ by the standard truncated-factorization semantics of interventions. 
\end{proof}

\subsection{Remark: Scope and Provenance}
The intervention operator realized as circuit surgery follows our companion work on \emph{Entanglement as Super-Confounding}~\cite{kang:2025:esc-arxiv}, where the operator is introduced and motivated conceptually.  Here we instantiate it within an SCM-to-circuit compiler and provide the above correctness guarantee for the compiled circuit family used in the main text.
We focus on functional correctness of compiled $\mathcal{DO}$-semantics on hardware; a detailed noise analysis is beyond the present scope.

%%%%%%%%%%%%%%%%%%%%%%%%%%%%%%%%%%%%%%%%%%%%%%%%%%%%%
\section{Compositional Semantics for \(\mathcal{DO}\), Conditioning, and Marginalization}
\label{app:compositional}

\noindent\textbf{Theorem (Compositional soundness).}\label{app:thm-soundness}
Let \(\mathcal{C}(G,\Theta)\) be the compiled circuit family used in the paper.
Then the following circuit primitives realize the corresponding SCM operations exactly:
(i) \emph{Interventions} \(\mathcal{DO}(S{=}s)\): circuit surgery (cut incoming controls to \(S\), prepare \(\ket{s}\));
(ii) \emph{Conditioning} on \(Z{=}z\): measure \(Z\) in the computational basis and post–select the branch \(z\);
(iii) \emph{Marginalization} over \(W\): classically mix measurement branches of \(W\).
For any expression formed by composing \(\mathcal{DO}\), conditioning, and marginalization,
\[
P^{\mathrm{circ}}(\cdot)\;=\;P^{\mathrm{SCM}}(\cdot).
\]
\emph{Proof sketch.} Item (i) is Proposition~\ref{prop:obs-do-correctness}(2). For (ii), post–selection implements restriction to the \(Z{=}z\) block (block–diagonal lemma), yielding \(P(\cdot\mid Z{=}z)\). For (iii), linearity of trace gives convex mixing over measured outcomes, which equals classical summation. Closure under sequential/parallel composition yields the claim. \(\square\)

\medskip
\noindent\textbf{Corollary (Backdoor adjustment via compiled circuit).}\label{cor:backdoor_adjustment}
For the Simpson DAG with confounder \(Z\),
\begin{align*}
    P^{\mathrm{circ}}(y\mid \mathcal{DO}(x)) &= \sum_{z} P(y\mid x,z)\,P(z) \\
    &= P^{\mathrm{SCM}}(y\mid \mathcal{DO}(x)).
\end{align*}
\emph{Proof.} Measure \(Z\), evaluate \(P(y\mid x,z)\) on each branch, and classically mix with weights \(P(z)\). Apply the theorem. \(\square\)

%%%%%%%%%%%%%%%%%%%%%%%%%%%%%%%%%%%%%%%%%%%%%%%%%%%%%
\section{Compilation Recipe: From Conditionals to \(\mathrm{CRY}\) Angles}
\label{app:compile_recipe}

\noindent The tables in the main appendix list the concrete angles actually used. For completeness, we record the generic mapping used by our compiler.

\paragraph{Single–parent arrow \(A\!\to\!B\).}
Initialize \(B\) in \(\lvert 0\rangle\).
If \(P(B{=}1\mid A{=}1)=p_{1}\), apply a single \(\mathrm{CRY}_{A\to B}(\theta_{1})\) controlled on \(A{=}1\) with
\[
\theta_{1}\;=\;2\arcsin\!\sqrt{p_{1}},\qquad
\text{so that}\quad \sin^2\!\bigl(\tfrac{\theta_{1}}{2}\bigr)=p_{1}.
\]
If you need a link conditioned on \(A{=}0\), wrap control by \(X\) on \(A\): \(X\)–\(\mathrm{CRY}\)–\(X\).

\paragraph{Two–parent arrow \((A,B)\!\to\!C\).}
For each parent pattern \((a,b)\in\{0,1\}^2\), if the desired conditional is \(P(C{=}1\mid a,b)=p_{ab}\), apply a \emph{mutually exclusive} doubly–controlled rotation \(\mathrm{CRY}_{(a,b)\to C}(\theta_{ab})\) that fires only on \((A{=}a,B{=}b)\), with
\[
\theta_{ab}\;=\;2\arcsin\!\sqrt{p_{ab}}.
\]
Because the controls are disjoint, only one gate can be active on any branch; the order of these gates is therefore immaterial, and the resulting probability on that branch equals \(p_{ab}\).

\paragraph{Multi–parent generalization.}
For \(k\) parents, decompose into \(2^{k}\) mutually exclusive control patterns and set \(\theta_{\mathbf{a}}=2\arcsin\sqrt{p(C{=}1\mid \mathbf{a})}\) for each pattern \(\mathbf{a}\).
Controls on logical \(0\) are implemented by \(X\)–wrapping the corresponding qubit.

\medskip
\noindent\emph{Remark.} This ``exclusive–control'' construction is the one instantiated by the angle tables. When hardware connectivity requires it, standard decompositions (e.g., ancilla–assisted Toffoli ladders) are used without altering the angles; exclusivity of controls is preserved by construction.

\end{document}